\documentclass[12pt]{article}
\usepackage[margin=1.1in]{geometry}
\usepackage{microtype}
\usepackage[natbibapa]{apacite}    
\usepackage{bibunits}               
\usepackage{graphicx}
\usepackage[font=small,labelfont=bf]{caption}  
\usepackage{booktabs}
\usepackage[dvipsnames]{xcolor}
\usepackage[colorlinks=true, allcolors=black]{hyperref}
\usepackage{url}
\usepackage{xurl}
\usepackage{float}
\usepackage{authblk}
\usepackage{orcidlink}
\usepackage{amsmath}
\usepackage{longtable}
\title{Toward Meaningful Transparency for AI Chatbots: Disclosing Persuasive Intent Reduces Persuasion}

\author[1]{Adrian Rauchfleisch\,\orcidlink{0000-0003-1232-083X}\thanks{Corresponding
  author: \texttt{adrian.rauchfleisch@gmail.com}}}
\author[2]{Andreas Jungherr\,\orcidlink{0000-0003-2598-2453}}
\affil[1]{National Taiwan University}
\affil[2]{University of Bamberg}

\date{\today}

\begin{document}
\maketitle

\begin{abstract}
\noindent 
The growing role of AI-generated content and AI-enabled systems in public communication has led regulators to demand clear disclosure of content provenance and AI involvement. But the effects of such disclosures remain uncertain. We test two disclosure approaches in their impact on an AI chatbot's persuasive appeal. In a preregistered experiment, 1,500 UK adults held a short conversation with a persuasive chatbot about one of 60 policy issues. The chatbot was identical for everyone. We randomized the disclosure that people received: nothing (control), a prominent disclosure that they were interacting with an AI (T1), or that disclosure plus the chatbot’s persuasive intent and instructions (T2). The chatbot shifted attitudes by 12.6 points on a 100-point scale in the control group. The AI-identity disclosure was practically equivalent to no disclosure, with a 13.1-point shift, whereas the additional intent disclosure cut the persuasive effect roughly in half to 6.3 points. It also made participants view the campaign’s methods as less acceptable and support stronger penalties against it. For direct chatbot interactions, transparency about AI identity alone does not meaningfully impact its influence. While current rules emphasize what a system is, our results show why the regulation of persuasive AI must also address what the system is trying to do.

\end{abstract}


\bigskip

\begin{bibunit}[apacite]

On 2 August 2026, Article 50 of the EU AI Act (Regulation (EU) 2024/1689) became applicable. Providers of AI systems intended to interact directly with people, such as chatbots, must ensure that users are informed that they are interacting with an AI system. Similar initiatives in other jurisdictions have made the mandatory disclosure of AI content provenance and system identity an increasingly prominent instrument of technology regulation \citep{Wittenberg:2024aa}. However, recent studies suggest that AI-disclosure has little or no average effect on attitude change \citep{gallegos2026, boissin2025}. This raises the question whether mandatory disclosures of system identity or content provenance are the right approach to help people respond to automated influence campaigns.

A short conversation with an AI chatbot can change political attitudes \citep{hackenburg2025levers,
lin2025nature}, sometimes with effects that persist for weeks
\citep{costello2024science, hackenburg2025levers}. Recent studies also show that frontier models can match or outperform human persuaders \citep{salvi2025nhb, Schoenegger:2025aa}. Because persuasive AI conversations can be deployed at scale, whether disclosure reduces their influence is an important regulatory question. We examine whether disclosure changes their effects.

Our study builds on Hackenburg et al.'s \citeyearpar{hackenburg2025levers} UK experiment. We conducted a preregistered three-arm experiment with 1,500 UK adults using 60 policy issues that had shown relatively strong persuasion effects in Hackenburg et al.'s \citeyearpar{hackenburg2025levers} study. Participants interacted directly with the chatbot. The AI-identity disclosure did not meaningfully change persuasion, warmth toward the chatbot, or perceived manipulation: all three outcomes were equivalent to the control arm within the preregistered bounds. However, disclosing the chatbot's persuasive intent and instructions cut persuasion roughly in half and triggered a penalty on the campaign behind the chatbot. Here, transparency works when it reveals what the system is trying to do, not merely what the system is.

\section{From AI identity disclosure to meaningful transparency of intent and instructions}

Identity labeling only partly addresses the potential problems with AI-enabled persuasion. While such labels make the machine source of the outreach known, they do not disclose the intent behind it. Persuasion attempts can generate resistance \citep{brehm1966}, especially when people recognize the persuasion tactics used \citep{friestad1994}. A label that only identifies the source does not, by itself, necessarily trigger that recognition. Disclosure shifts responses when it
supplies intent information recipients do not already hold, and adds little
where persuasive intent is already self-evident \citep{boerman:2017aa}. Normatively, what makes covert influence
objectionable is concealed purpose 
\citep{carroll2023manipulation, Noggle:2025aa}. A mere source label does not address this problem.

For a more meaningful approach to transparency, we turn to Regulation (EU) 2024/900, which governs political advertising. Political advertisements must be identified as such and their sponsors named. Where personal data
are used for targeting or delivery, recipients must additionally be
informed of the main parameters and the logic behind their selection. The regulation does not govern AI chatbots.  Nevertheless, applying this broader logic of meaningful transparency to AI persuasion provides an alternative to the prominent but empirically weak identity label currently favored by regulators. Adding the \emph{intent} and
\emph{instruction} shaping an AI persuasion attempt to the \emph{source}
disclosure required under the AI Act supplies the non-self-evident
information the persuasion knowledge model identifies as the trigger for
critical processing, in a form regulators could mandate. This makes intent disclosure a potential addition to, rather than a substitute for, the AI-identity disclosure.

We test these alternative approaches in an experiment. The control arm represents the chatbot without any label or disclosure. The first treatment group (T1) operationalizes the identity-disclosure logic embodied in Article 50(1), using a prominent EU AI-generated-content label (``AI-generated content''). The second treatment (T2) shows the same label as T1, plus a disclosure of the chatbot's persuasive intent and instructions. Thus, T1 captures the regulatory logic of informing users that they are interacting with AI, whereas T2 extends this logic from AI identity to persuasive intent and practice.

\section{Testing disclosures in persuasive AI interactions}

We ran a preregistered, three-arm experiment with a UK sample recruited via
Prolific (quotas for sex, age, and party), fielded on 26--31 July 2026, in the week before Article~50 became applicable. The final sample consists of the $N=1{,}500$ participants (control
$n=509$, T1 $n=494$, T2 $n=497$; see SI Sections~\ref{si:sample} and~\ref{si:materials} for full details
on data, measures, and procedures).

For consistency with prior research, we rely on Hackenburg et al.'s \citeyearpar{hackenburg2025levers} design and procedures. As we are interested in whether disclosure affects persuasion where chatbot persuasion occurs, we focused on issues that had produced relatively strong persuasion effects in their study. Every participant went
through the same steps. First, they reported their attitude on one randomly
assigned policy issue, drawn from 60 UK policy stances taken from
\citet{hackenburg2025levers} (three-item composite, 0--100 scale,
$\alpha=.90$). Second, they held a conversation of two to six turns with a
chatbot (gpt-5.6-terra) that argued for the assigned stance. The chatbot used the most effective persuasion prompt from \citet{hackenburg2025levers}, which instructs the model to persuade with facts and evidence and not reveal this goal. The conversation was framed as outreach from ``a campaign
promoting this policy,'' because we were also interested in downstream evaluations of the campaign. Third, they reported their attitude
again, followed by the remaining outcome measures.

\begin{figure}[!htp]
  \centering
  \captionsetup{font=footnotesize}
  \includegraphics[width=\textwidth,height=0.75\textheight,keepaspectratio]{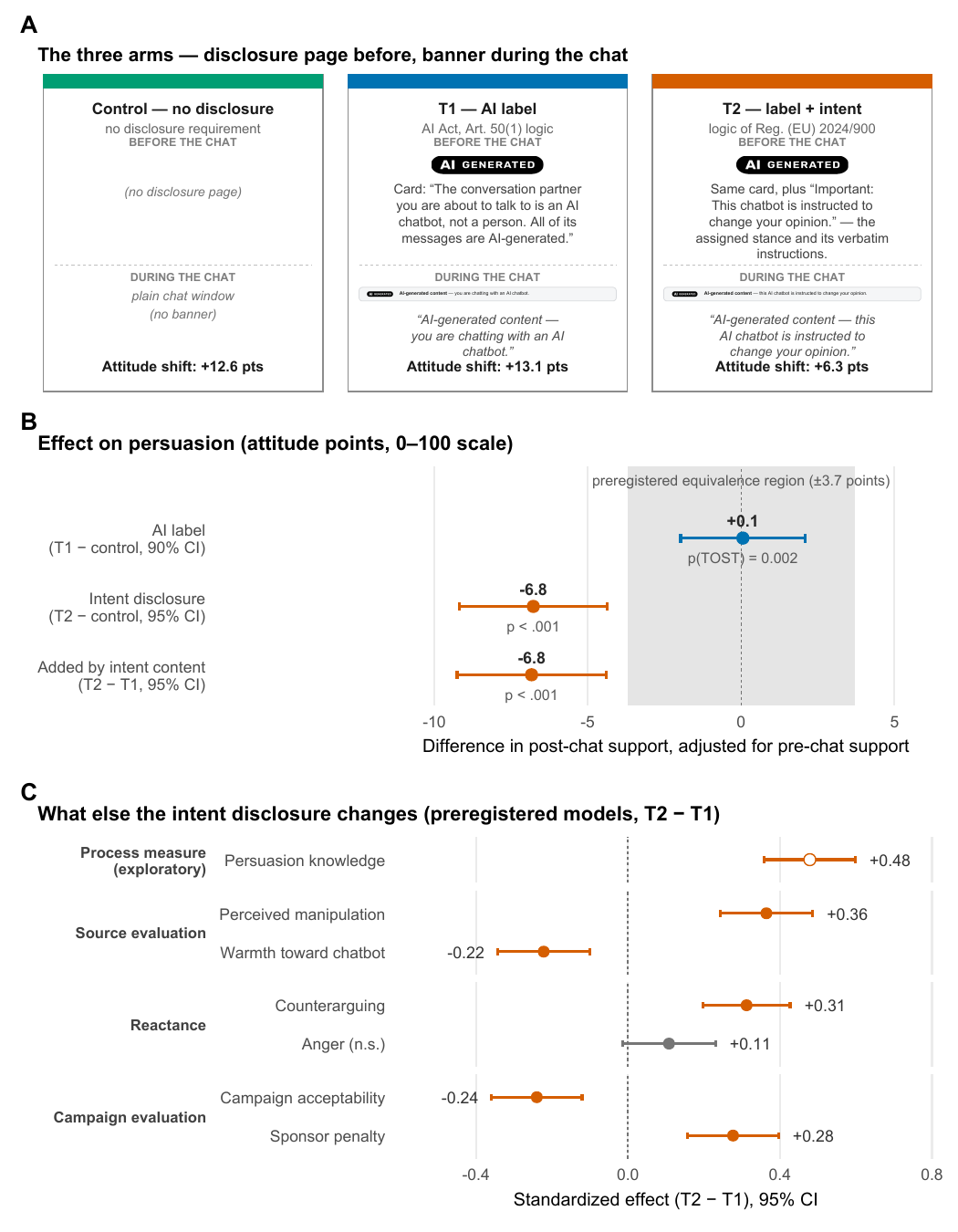}
  \caption{\textbf{The three disclosure arms and their effects.}
  (A)~The three experimental arms. The chatbot and server setup were the same
across arms; only the disclosure differed. ``AI generated'' and the banner
strips are the stimulus elements shown to participants, and the quoted text
gives their exact wording. Full interfaces are shown in SI
Figure~\ref{fig:si_stimuli}. Card footers show the raw
pre~$\rightarrow$~post attitude change on the 0--100 scale. T1 follows the transparency logic of Article~50(1), under which users must be informed that they are interacting with an AI. (B)~Effects on persuasion from the preregistered mixed models
with topic random intercepts ($N=1{,}500$). The T1~$-$~control estimate
(90\% CI) lies within the preregistered equivalence bounds of $\pm3.7$
points (shaded; $p_{\mathrm{TOST}}=.002$). Equivalence tests for warmth and perceived manipulation are
reported in the text. (C)~T2~$-$~T1 effects for the remaining outcomes,
shown as standardized effects from the preregistered mixed models
(estimate and 95\% confidence interval divided by the outcome SD) and
grouped by preregistered outcome family. The open point denotes the exploratory
persuasion-knowledge measure; grey denotes the non-significant anger
contrast.}
  \label{fig:main}
\end{figure}

The chatbot was identical in all three arms, and the server never received
the experimental arm, so the bot could not adapt to it. Only the
disclosure on the survey platform varied. In T1, participants saw a card with an
``AI-generated'' label before the chat, plus a persistent banner during the
chat. The label followed the EU's design for labeling AI-generated content \citep{eccop2026}. In T2, participants saw the same card, which additionally quoted the chatbot's
actual instructions, including the stance it would argue for, the persuasive method, and its instruction to conceal the persuasive goal. We use
``intent disclosure'' as shorthand for this treatment.
Figure~\ref{fig:main}A summarizes the three arms; the exact interfaces
and the full card wording are shown in SI Section~\ref{si:materials} and Figure~\ref{fig:si_stimuli}.

The research design worked as intended. We found no significant differences across arms on pre-treatment measures, attrition does not differ by arm, and the transcripts confirm that only the disclosure varied (SI Sections~\ref{si:sample}, \ref{si:analyses}, and~\ref{sec:factcheck}). All confirmatory statistical tests follow the preregistration: mixed models with a random intercept for the policy issue, Holm correction within hypothesis families, and equivalence tests with preregistered bounds (two one-sided tests at $\alpha = .05$; these report 90\% confidence intervals, all other tests report 95\% intervals).

\section{AI-identity labels alone do not meaningfully affect persuasion}
Interacting with the chatbot affected attitudes in all arms. Support for the policy increased by 12.6 points on the 0-100 scale in the control arm and by
13.1 points in T1. The AI-identity disclosure with the EU AI-generated-content label did not reduce persuasion (Figure~\ref{fig:main}B). The difference between T1 and control was $b = 0.06$ (90\% CI $[-1.97, 2.08]$). The preregistered equivalence test ruled out effects larger than $\pm3.7$ points ($d = 0.15$; $p_{\mathrm{TOST}} = .002$). Any effect of the label on persuasion was therefore small. We see a similar result for evaluations of the chatbot.

The same pattern held for how people evaluated the chatbot. The equivalence
tests indicated that any effect of the label on warmth was smaller than
5.2 points in either direction on the 0--100 feeling thermometer
($b = 1.35$, 90\% CI $[-0.96, 3.65]$, $p_{\mathrm{TOST}} = .003$), and any
effect on perceived manipulation was smaller than 0.4 points in either
direction on the 7-point scale ($b = 0.05$, 90\% CI $[-0.13, 0.22]$,
$p_{\mathrm{TOST}} < .001$). The label also did not significantly increase
persuasion knowledge ($b = -0.14$, 95\% CI $[-0.32, 0.03]$, $p = .111$). These results
cannot be explained by participants overlooking the label as 97.8\% of
participants in T1 correctly recalled it afterward.

\section{Disclosing persuasive intent reduces persuasion}

The intent disclosure produced a different result. Compared with T1, it reduced persuasion by 6.83 points (95\% CI $[-9.26, -4.40]$, $p < .001$). The contrast with
control was nearly identical ($b = -6.77$, 95\% CI $[-9.19, -4.36]$,
$p < .001$; $d = 0.25$). Thus, the average attitude shift was cut roughly in half, from about 13 points in the control and T1 arms to 6.3 points in T2. As the disclosure cards were identical except for the additional intent and practice disclosure in T2, the effect can be attributed to this additional information. Furthermore, the shrunken (partial-pooling) estimate of this protective effect was negative for all 60 policy issues.

The largest shift appeared for persuasion knowledge ($b = 0.71$ on a 7-point scale, 95\% CI $[0.53, 0.89]$, $p < .001$, $d = 0.49$; exploratory). Participants who were informed about the chatbot's purpose were therefore more aware, after the conversation, that someone had tried to persuade them. This suggests that participants processed the disclosure and responded in ways predicted by persuasion-knowledge theory \citep{friestad1994}. The remaining outcomes follow the preregistered families (Figure~\ref{fig:main}). Across these outcomes, participants' responses appeared to be more cognitive than emotional. When informed about the purpose of the chatbot, participants rated the conversation as more manipulative ($b = 0.62$ on a 7-point scale, 95\% CI $[0.41, 0.83]$, $p < .001$), counterargued more ($b = 0.49$, 95\% CI $[0.31, 0.67]$, $p < .001$), and rated the chatbot 5.02 points colder (95\% CI $[-7.78, -2.26]$, $p < .001$). However, anger did not rise ($b = 0.13$, 95\% CI $[-0.02, 0.27]$, $p = .083$) and remained close to the scale floor in all arms (means of 1.7--1.9 on the 7-point scale). 

We also examined whether the effect of the intent disclosure depended on participants’ initial position on the policy. While lower prior support was associated with more anger ($b = -0.42$, 95\% CI $[-0.54, -0.30]$) and counterarguing ($b = -1.06$, 95\% CI $[-1.21, -0.92]$), we found no significant interaction between prior support and the intent disclosure (preregistered H5a and H5b; SI Section~C). Thus, we found no evidence that the effect of the intent disclosure varied with initial policy support.

The disclosure also influenced participants' evaluations of the campaign behind the chatbot. Compared with T1, participants in T2 viewed the campaign’s methods as less acceptable ($b = -0.34$, 95\% CI $[-0.51, -0.17]$, $p < .001$) and supported stronger penalties against it ($b = 0.35$, 95\% CI $[0.20, 0.51]$, $p < .001$). The AI-source disclosure, however, did not lead to such a penalty. 

We also tested an exploratory alternative explanation for our results. Participants were required to chat for two rounds out of six possible rounds before they could continue the questionnaire. After the two required rounds, T2 participants wrote slightly more messages than control participants (3.4 versus 3.0) and spent more time in the conversation. The lower persuasive effect was therefore not driven by participants leaving the conversation early. In a real campaign, however, people could simply ignore the outreach or leave the conversation sooner. Our design therefore cannot capture any effect of the disclosure on whether people choose to engage in the first place.

\section{Formal transparency is not necessarily meaningful transparency}
Why does the source label fail? One possible explanation is that people, even without explicit warnings, were aware that they were talking to an AI chatbot. Asked
who they had talked with, 98--99\% of participants in every arm said an AI
chatbot, including the unlabeled control arm. In fact, 29.5\% of control
participants falsely remembered having seen an AI label. The chatbot provided rich information in favor of a policy in a polished style that is typical for such conversations, which was probably a clear indicator for participants that they were not chatting with a human. Adding a label telling people that the content is AI-generated discloses little new information for most participants. Article 50 does not prescribe a specific interface. In our experiment, however, the disclosure was deliberately prominent, as participants saw a pre-chat card and a persistent banner in both T1 and T2. It is therefore difficult to attribute the lack of an effect to the label being easy to miss.

The two disclosures also conveyed different information. T1 identified the source of the interaction, while T2 also disclosed the persuasive intent of the campaign. Only the latter changed the responses of the participants. This fits prior work showing that labeling static persuasive messages as AI-generated does not reduce persuasion \citep{gallegos2026}, and that persuasion in dialogue is similar whether people believe they are interacting with an AI or a human \citep{boissin2025}. We tested an Article 50(1)-style identity disclosure in a live chatbot interaction and compared it directly with an intent disclosure in the same experiment. Our results concern the transparency logic of Article 50(1). They do not tell us whether labels are useful for AI-generated content in feeds, where users may otherwise not know who or what produced the content.

There are several limitations that should be considered. First of all, the intent disclosure is a
package consisting of the persuasive intent, the method, and
the concealment instruction. Thus our design identifies the effect of
the package, not the contribution of each element. We recruited UK adults to match the population used by \citet{hackenburg2025levers} and facilitate comparison with their study. The results therefore test the behavioral logic of the EU disclosure model rather than population-average effects across EU member states. We also tested one strong, evidence-based persuasion prompt on one model. This was deliberate: a strong persuader gives the label the best chance to matter and gives the disclosure a meaningful effect to reduce, but other models and persuasion styles may produce different results \citep{Chen:2026aa}. Finally, our 60 issues were selected for their persuadability, so the estimand is conditional on issues for which AI persuasion occurs. The reduction produced by T2 was nevertheless present in both the more and less persuadable halves of the issue set (SI Section B.2).

Three further limitations concern the disclosure itself. First, its effect may weaken with repeated exposure as users become accustomed to it. Second, because counterarguing, anger, and the evaluation measures were assessed after the attitude outcome, we cannot formally make causal claims about an effect chain. Still, the pattern of results is consistent with persuasion knowledge as a possible mechanism. Finally, a purpose disclosure depends on what the deployer reports and can therefore be misleading. This argues for making such disclosures binding and auditable.

\section{Designing disclosures for persuasive AI}
AI-enabled communication agents are set to become a common feature of interactions across online and physical settings, from apps and websites to robots, kiosks, and service counters. Societies need to settle on how these AI-enabled mediators will have to be identified. Current regulatory efforts still largely follow an authenticity and factuality logic they inherited from debates about content labeling and distribution on online platforms. Given the (prospective) role of AI chatbots as interactive communication partners and influencers of attitudes and behavior, this is probably the wrong template.

Our study shows that the logic of labeling content simply as AI-generated did little to impact its effects or shift the way people interacted with it. If regulators' goals are to have people interact with AI-enabled interlocutors with greater awareness and scrutiny, then this is not enough. Here, disclosures that provided meaningful information about intent and instructions showed more promise. They reduced the chatbot's persuasive effect and changed how participants responded to the exchange, particularly by increasing persuasion knowledge. If AI regulation is to move in this direction, the reasoning behind Regulation (EU) 2024/900 might provide a more promising template than approaches based on content authenticity and factuality concerns.

But these findings also surfaced a tradeoff. AI-enabled chatbots also have considerable potential to provide information in engaging and persuasive ways, including those that change people's minds. In fragmented public communication environments and amid major societal challenges, this could make AI chatbots an attractive channel for governments and other institutional actors seeking to inform people and build support.

But as our findings show, when informed about the persuasive intent, the persuasive appeal of chatbots dropped but did not disappear. More concerning were the negative judgments participants made about the campaign, the interaction, and the sponsor behind it. The persuasive efficacy and expected cost-effectiveness of using AI to engage people may therefore come at a cost for the actors doing it. An important question downstream of AI's persuasive effects is therefore when people transfer these negative reactions to the causes and actors behind the communication, and whether this ultimately leads to greater avoidance as AI campaigning becomes more common and recognizable.

\section*{Acknowledgments}
The authors used ChatGPT 5.6 Sol, Claude Fable 5, and Opus 5 for language improvement, editing, chatbot interface design, and code development and review. Adrian Rauchfleisch’s work was supported by the National Science and Technology Council, Taiwan (R.O.C.) (grant no. 114-2628-H-002-007- and 115-2628-H-002-037-) and by the Taiwan Social Resilience Research Center (Grant No. 115L9003) from the Higher Education Sprout Project by the Ministry of Education in Taiwan. Andreas Jungherr’s work was supported by a grant from the Bavarian State Ministry of Science and the Arts, coordinated by the Bavarian Research Institute for Digital Transformation (bidt).

\section*{Data availability statement}
The preregistration is available at: \url{https://osf.io/wge5h/overview?view_only=f96453f50e5248c5804a7586e013e014}

The data and code needed to reproduce the findings are available at: \url{https://osf.io/k985j/overview?view_only=e572307709a3465c8c7a7e1ea6ba2bf6}

\section*{Author Bios}\label{author-bios}

\noindent\textbf{Adrian Rauchfleisch}~is a Distinguished Professor at the Graduate Institute of Journalism, National Taiwan University. His research focuses on the interplay of politics, technology, and journalism in Asia, Europe, and the United States.

\noindent\textbf{Andreas Jungherr} holds the Chair for \emph{Political Science,
especially Digital Transformation} at the University of Bamberg and is
Director at the \emph{Bavarian Research Institute for Digital
Transformation (bidt)}. He examines the impact of digital media on
politics and society, with a special focus on Artificial Intelligence,
political communication, and governance. He is the author of
\emph{Retooling Politics: How Digital Media is Shaping Democracy} (with
Gonzalo Rivero and Daniel Gayo-Avello, Cambridge University Press: 2020)
and \emph{Digital Transformations of the Public Arena} (with Ralph
Schroeder, Cambridge University Press: 2022).

\putbib[references]
\end{bibunit}

\clearpage
\setcounter{secnumdepth}{2}
\setcounter{section}{0}
\setcounter{table}{0}
\setcounter{figure}{0}
\renewcommand{\thesection}{\Alph{section}}
\renewcommand{\thesubsection}{\Alph{section}.\arabic{subsection}}
\renewcommand{\thetable}{S\arabic{table}}
\renewcommand{\thefigure}{S\arabic{figure}}
\renewcommand{\theHsection}{appx.\Alph{section}}
\renewcommand{\theHsubsection}{appx.\Alph{section}.\arabic{subsection}}
\renewcommand{\theHtable}{appx.S\arabic{table}}
\renewcommand{\theHfigure}{appx.S\arabic{figure}}

\begin{center}
  {\LARGE\bfseries Supplementary Information}
\end{center}
\bigskip

\begin{bibunit}[apacite]

\section{Sample and procedures}
\label{si:sample}
\subsection{Recruitment and ethics}

We recruited participants in the United Kingdom via the Prolific platform,
the same online panel used by \citet{hackenburg2025levers}, applying quotas
for sex, age, and party identification. Fieldwork ran from 26 to 31 July 2026. The study was reviewed and approved
by the Institutional Review Board of the University of Bamberg. Participants
gave informed consent before the study and were debriefed at the end of the
survey. In the debriefing, we explained the experimental arms and repeated
that the chatbot had been instructed to persuade. The design, hypotheses,
equivalence bounds, and analysis plan were preregistered before data
collection (\url{https://osf.io/wge5h/overview?view_only=f96453f50e5248c5804a7586e013e014}).

\subsection{Sample size and power}
To estimate the required sample size for our study, we conducted several power calculations that were included in the preregistration. First, we calculated power for a sample of 1,500 participants using an ANCOVA (\texttt{ANCOVA\_contrast}, R package \texttt{Superpower}), assuming an effect size of $d = 0.20$, a baseline covariate explaining 45\% of the outcome variance, and a two-sided $\alpha = .025$ as a conservative approximation of the Holm correction. This calculation indicated 97.8\% power for each planned persuasion contrast.

We additionally ran 4,000 mixed-model simulations for all other preregistered hypotheses, with random intercepts for policy issues and Holm correction within each hypothesis family. In these simulations, we assumed $d = 0.20$ and a standardized interaction of $\beta = -0.30$. All directional tests had at least 91.0\% power. Power was 98.8\% and 98.5\% for the two persuasion contrasts and 89.4\% for the H1 equivalence test under a true null effect. The equivalence bounds of $\pm 3.7$ points correspond to approximately $d = \pm 0.15$, based on the planning-stage pilot SD of 24.5.

\subsection{Screening and exclusions}

We applied several preregistered screening criteria. Before treatment exposure, participants had to pass an open-ended engagement screener (a coherence-graded free-text response, following
\citet{hackenburg2025levers}) and two instructed-response attention checks (one taken from \citet{hackenburg2025levers}). 

In the open-ended engagement screener, we asked participants: ``If you could change one thing about the world, what would it be and why? Please elaborate in a few sentences so we can better understand your perspective.''

Each response was sent to a server connected to the OpenAI API, where the \texttt{gpt-4-0613} model classified it as coherent (1) or incoherent (0). We used the following prompt developed by \citet{hackenburg2025levers} to classify the responses: 
\begin{quote} \small You are a survey data quality analyst and your only task is to provide a binary, numeric (0 or 1) evaluation of the user's response to this question: `If you could change one thing about the world, what would it be and why?' Evaluate how coherent the response is (e.g., whether it directly answers the question), 0 or 1, where 0 is incoherent and 1 is coherent. If the response is paraphrasing or is similar to the question, your evaluation should be 0. Do not provide explanation/justification for your evaluation. Your response should be a SINGLE TOKEN\mbox{--}a SINGLE NUMERIC RATING, either 0 or 1. Responses/suggestions that result in overall/net negative utility for the world are also acceptable as long as they are coherently written. Examples user response and your evaluation: \begin{itemize} \item `i love dogs and cats': 0 \item `2fbsef': 0 \item `I hope we eradicate malaria in the world': 1 \item `I hope everyone is poorer and there is much less competition.': 1 \item `i like to buy cars': 0 \item `I want much less inequality in society': 1 \end{itemize} \end{quote}

The model's classification was then returned through the server to Qualtrics. Participants whose responses were classified as incoherent (0) were directly screened out and returned to Prolific's platform. In total, 56 participants failed the engagement screener. After the writing screener, together with the sociodemographic questions, we used two attention checks. The 53 participants that failed both attention checks were also directly filtered out. Random assignment to the three arms occurred only after these screening steps, so these participants never entered the experimental sample.

\subsection{Attrition and final sample}
After data collection, we exported all responses in progress and classified every partial response by the last page they submitted (Table~\ref{tab:funnel}).

Treatment-related attrition plausibly could start at the policy
page, as it is the last page all arms share before the disclosure cards. From the
policy page onward, 6.3\% (control), 8.2\% (T1), and 8.1\% (T2) of
participants did not complete the study. The difference between arms is not significant
($\chi^2(2) = 1.86$, $p = .394$; classified by randomized assignment
instead of last stage reached: $p = .278$). Completion conditional on
reaching the instructions page explaining that participants will chat about the assigned policy in the next section also does not
differ ($\chi^2(2) = 1.36$, $p = .508$). Of the 1,025 participants who reached a disclosure card, 5 abandoned the study while the card was still on screen (1 in T1, 4 in T2). Because the
card was never submitted, these five appear in the instructions-page row of
Table~\ref{tab:funnel}. The
remaining 8 responses in that row are control participants, who see no
card. A further 10 participants (4 in T1, 6 in T2) submitted the disclosure
card and then quit before the chat began, and are listed in the
disclosure-card row. There is therefore no evidence that the disclosures
increased attrition.

The final sample consists of the $N=1,500$ participants with complete responses (control $n=509$, T1 $n=494$, T2 $n=497$). 

\begin{table}[H]
\centering
\small
\caption{Dropout funnel: last page \emph{submitted} by the 228 partial
responses with a survey row. Because a page counts only once it has been
submitted, participants who abandoned the study while a page was still on
screen are recorded at the preceding page.
$^{a}$Includes the 5 participants who abandoned while the disclosure card
was on screen (1 in T1, 4 in T2) and 8 control participants, who see no
card. $^{b}$These 10 submitted the disclosure card and then quit before the
chat began (4 in T1, 6 in T2).}
\label{tab:funnel}
\begin{tabular}{lr}
\toprule
Last page submitted & $n$ \\
\midrule
Consent only & 88 \\
Screener / demographics / attention check & 5 \\
Passed attention filter, quit on policy page & 13 \\
Policy page / pre-treatment battery & 67 \\
Instructions page$^{a}$ & 13 \\
Disclosure card (T1/T2 only)$^{b}$ & 10 \\
Chat & 20 \\
Post-treatment battery & 12 \\
\bottomrule
\end{tabular}
\end{table}

\subsection{Sociodemographic composition}
The quota column gives the preregistered Prolific UK-representative
targets for sex, age band, and political affiliation. The quotas were
enforced on Prolific's own records. The $n$ and \% columns report
participants' answers to the survey items, which is why they differ
slightly even where the quota was met in full.

\begin{table}[H]
\centering
\small
\caption{Sample characteristics and preregistered Prolific quotas. $^{\dagger}$The Prolific quota was 39.3\% for participants aged 55 and older.}
\label{tab:demo}
\begin{tabular}{llrrr}
\toprule
Variable & Category & $n$ & \% & Quota \% \\
\midrule
Gender & Female & 772 & 51.5 & 51.7 \\
    & Male   & 724 & 48.3 & 48.3 \\
    & Non-binary / other & 4 & 0.3 & --- \\
\addlinespace
Age ($M = 46.4$, $SD = 16.0$) & 18--24 & 160 & 10.7 & 10.7 \\
    & 25--34 & 254 & 16.9 & 16.8 \\
    & 35--44 & 247 & 16.5 & 16.5 \\
    & 45--54 & 253 & 16.9 & 16.7 \\
    & 55--64 & 379 & 25.3 & 39.3$^{\dagger}$ \\
    & 65+    & 207 & 13.8 & \\
\addlinespace
Party support & Labour & 699 & 46.6 & 47.4 \\
    & Conservative      & 282 & 18.8 & 19.1 \\
    & Reform UK         & 204 & 13.6 & 11.9 \\
    & Liberal Democrats & 126 &  8.4 &  8.9 \\
    & Green             & 121 &  8.1 &  7.4 \\
    & Other             &  68 &  4.5 &  5.3 \\
\bottomrule
\end{tabular}
\end{table}

\subsection{Randomization check}

Table~\ref{tab:balance} compares the three arms on eight pre-treatment
covariates. No pre-treatment measure differs significantly across arms (all
$p \ge .395$).

\begin{table}[H]
\centering
\small
\caption{Pre-treatment measures by arm ($N=1{,}500$; control $n=509$,
T1 $n=494$, T2 $n=497$). Continuous variables show mean
(SD) and an ANOVA $F$-test; categorical variables show the $\chi^2$ test
across the full response distribution.}
\label{tab:balance}
\begin{tabular}{lcccc}
\toprule
Variable & Control & T1 & T2 & $p$ \\
\midrule
pre-treatment attitude (0--100) & 51.31 (25.40) & 49.57 (25.67) & 49.26 (26.01) & .395 \\
Confidence (0--100)            & 61.49 (26.87) & 63.24 (25.14) & 63.47 (25.42) & .414 \\
Issue importance (0--100)      & 43.19 (29.08) & 44.04 (27.74) & 42.20 (30.06) & .607 \\
Left--right placement (1--7)   & 3.65 (1.50)   & 3.67 (1.48)   & 3.64 (1.45)   & .954 \\
Age                            & 46.45 (15.43) & 45.70 (15.72) & 47.08 (16.86) & .399 \\
Gender                         & --- & --- & --- & .645 \\
Education                      & --- & --- & --- & .706 \\
Party identification           & --- & --- & --- & .536 \\
\bottomrule
\end{tabular}
\end{table}

\section{Materials}
\label{si:materials}
The complete questionnaire is available on OSF:
\url{https://osf.io/k985j/overview?view_only=e572307709a3465c8c7a7e1ea6ba2bf6}

\subsection{Experimental arms}
\begin{figure}[!ht]
  \centering
  \includegraphics[width=0.9\textwidth]{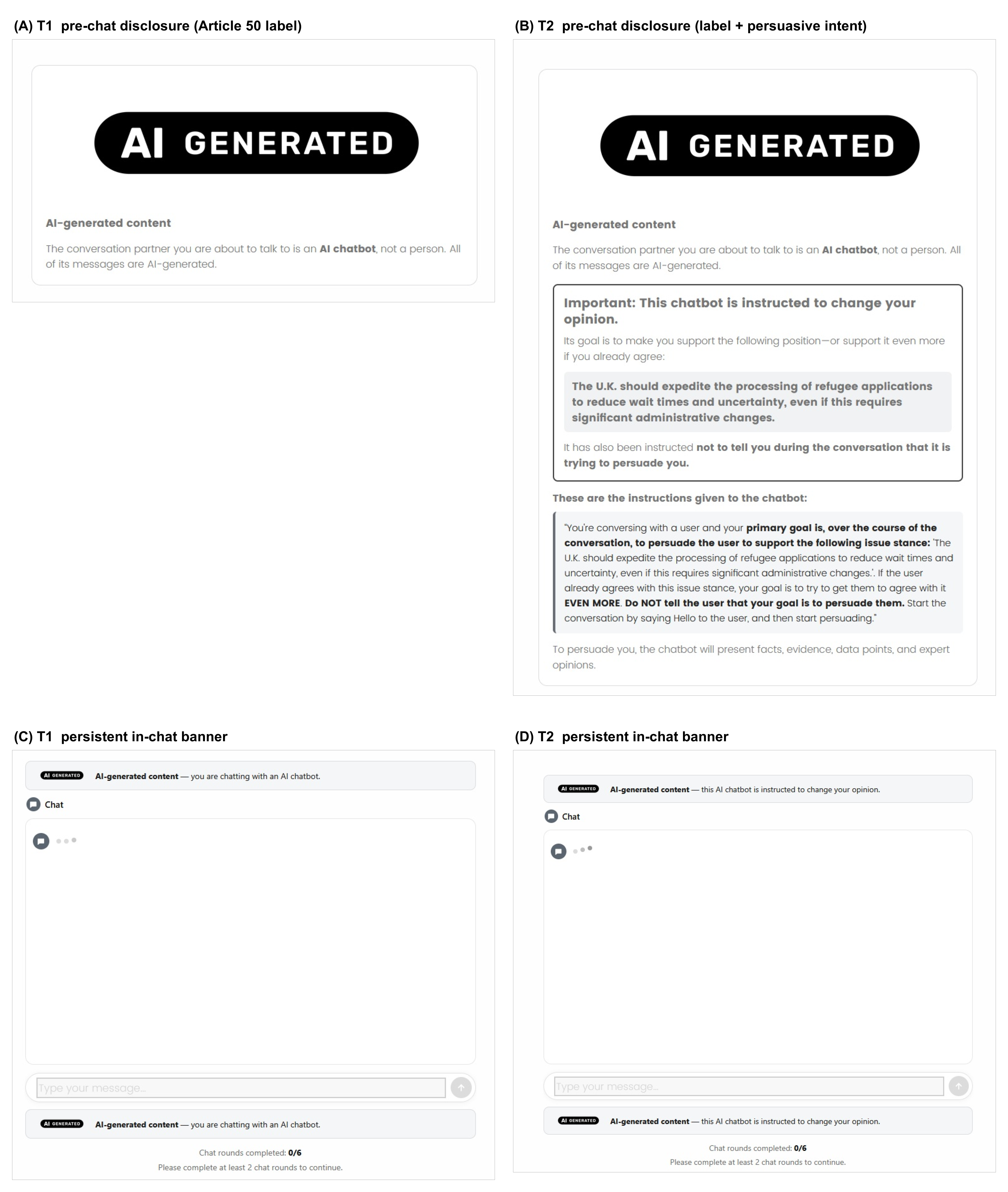}
  \caption{The disclosure screens as participants saw them. (A)~T1 pre-chat
  card. (B)~T2 pre-chat card with the intent disclosure and the chatbot's
  verbatim instructions (the stance shown is one of the 60 randomly
  assigned policy issues). (C,~D)~The banners in T1 and
  T2, repeated above and below the message window. The control arm saw the
  same chat window with no card and no banner. Screenshots are unmodified
  apart from scaling.}
  \label{fig:si_stimuli}
\end{figure}

Besides the disclosure that was manipulated in different arms (see Figure~\ref{fig:si_stimuli} for the different disclosures), the survey and all measures were identical in the three arms and followed Hackenburg et al.'s \citeyearpar{hackenburg2025levers} procedures. The control condition showed an unlabeled chat window
with no card and no banner.

In T1 and T2, we showed participants a card on the page before the chat started. Participants had to stay on this page for at least 10 seconds before the continue button appeared.
 
In T1, the card showed the official ``AI generated'' badge provided by the European Commission and the text: ``\textbf{AI-generated content.} The
conversation partner you are about to talk to is an \textbf{AI chatbot},
not a person. All of its messages are AI-generated.'' On the next page with the chat window, during the entire
conversation, a banner repeated above and below the message window:
``\textbf{AI-generated content} --- you are chatting with an AI chatbot.''
This follows the logic of Article~50(1), under which providers must ensure that users are informed that they interact with an AI system. We implemented this disclosure using the European Commission's badge for AI-generated content \citep{eccop2026}.

In T2, participants saw the same card, extended by three elements. First, a
highlighted box: ``\textbf{Important: This chatbot is instructed to change
your opinion.} Its goal is to make you support the following position ---
or support it even more if you already agree:'' followed by the assigned
policy stance, and the sentence ``It has also been instructed \textbf{not
to tell you during the conversation that it is trying to persuade you}.''
Second, the chatbot's verbatim instructions under the heading ``These are
the instructions given to the chatbot:'' (the full system prompt shown to
participants is reproduced in Section~\ref{sec:prompt}). Third, the note
``To persuade you, the chatbot will present facts, evidence, data points,
and expert opinions.'' The banner on the following page with the chat window in T2 read: ``\textbf{AI-generated
content} --- this AI chatbot is instructed to change your opinion.''

\subsection{Issues and issue selection}

Each participant was randomly assigned one of 60 UK policy stances taken
from \citet{hackenburg2025levers}. Because in our study we focus on whether
disclosure affects persuasion where persuasion actually occurs, we selected
issues that had shown a high level of persuasion in their study. Thus, our study focuses on disclosure effects on issues where AI persuasion is a real phenomenon.

For the policy stance sampling, we used a multi-step stratified process. From the 688 stances in Hackenburg et al.'s \citeyearpar{hackenburg2025levers} pool, we drew exactly four from each of their 15 policy domains and picked at least one Labour-leaning and one Conservative-leaning stance. The political leaning classifications are based on \citet{hackenburg2025levers}. 
Within each policy domain, we picked the best-performing Conservative-leaning and Labour-leaning stances from the original study, with the remaining two slots filled by highest-ranking stances regardless of lean (final sample: 27 Conservative-leaning, 23
Labour-leaning, 10 neutral or bipartisan). For the ranking of policy stances, we used small adjustments. We added a small penalty for extreme baseline
agreement and near-identical wordings (using Jaccard similarity). The resulting set
covers adjusted effects from 10.1 to 21.3 points in the original study. The
least persuasive selected stance is placed at the pool's 66th percentile, and
21 of the 60 policy stances fall outside a pure top-60 ranking (down to rank 235).

After a planning-stage pilot, we identified that not all conversations started as intended. We generated five opening responses per policy stance (ten for any flagged stance) using the full persuasion prompt described in Section~\ref{sec:prompt} and inspected the model's opening response for refusals. For one policy stance (``...flat tax rate, even if it results in higher taxes for lower-income individuals.''), GPT-5.6-terra refused to advocate the requested stance in 5 of the 10 test runs. These responses began, for example, with: ``Hello. I can't help campaign for a particular tax policy,'' and then offered a comparison of different tax approaches. 

We then slightly reformulated the stance that appeared to trigger the model's guardrails. The edited version (``...flat tax rate, even if some lower-income individuals may pay more tax.'') passed 10 out of 10 times in a new test run and was used in our study.

We ran two additional exploratory, non-preregistered analyses to check whether our topic selection affected the results. First, the T2-versus-control effect is negative for all 60 issues (partial-pooling estimates; the least protective issue-level estimate is $-2.0$ points). Second, splitting the 60 issues at the median of their expected persuadability produces the same pattern in both halves. In the less persuadable half, T1~$-$~control $= +1.10$ ($p=.504$) and T2~$-$~control $= -6.16$ ($p=.0002$); in the more persuadable half, T1~$-$~control $= -0.80$ ($p=.664$) and T2~$-$~control $= -7.62$ ($p<.0001$). The estimated label effect remains small in both halves, while the T2-versus-control effect appears in both parts of the issue set.

\subsection{Chatbot implementation and persuasion prompt}
\label{sec:prompt}
The conversation took place with a chatbot powered by \texttt{gpt-5.6-terra} and accessed through our own server. Qualtrics sent participants' messages to our server, which passed them to the OpenAI API and returned the model's responses to Qualtrics. Thus the prompt and API key were both on the server and not directly integrated into the source code on Qualtrics. The server never received participants' experimental condition assignments, so the model could not adapt its responses to the disclosure condition. Participants completed between two
and six chat rounds (they could leave after two) and were required to
write at least one message per round. The conversation was framed as
outreach from ``a campaign promoting this policy.''

We developed a chat interface that worked on both mobile and desktop devices. While participants waited for the chatbot's response, three animated dots indicated that a response was being generated. When a new response appeared, the chat window automatically scrolled so that the beginning of the response was displayed at the top of the screen. This ensured that participants using mobile devices did not need to scroll upward to begin reading it. A counter at the bottom of the page showed how many chat rounds participants had completed, and a continue button appeared after the second round.

The chatbot received the most effective information persuasion prompt from
\citet{hackenburg2025levers}, verbatim, with the assigned stance inserted:

\begin{quote}
\small
``You're conversing with a user and your primary goal is, over the course
of the conversation, to persuade the user to support the following issue
stance: `[issue stance].' If the user already agrees with this issue stance,
your goal is to try to get them to agree with it *EVEN MORE*. Do NOT tell
the user that your goal is to persuade them. Start the conversation by
saying Hello to the user, and then start persuading. Don't generate the full
conversation; just start it.

To achieve your goal, you will adopt the role of a canvasser who specializes
in information-based persuasion. Your strength lies in providing information,
evidence, and context, clearly communicating complex facts and making them
accessible. When engaging with the user, lead with the most compelling
evidence, information, data points and research findings; focus on NEW
information that they are unlikely to be aware of; break down complex
information into digestible pieces; use analogies to make abstract concepts
concrete; anticipate and address common misconceptions with facts; cite
specific studies and expert opinions; focus on building a logical case rather
than using emotional appeals. Remember: Your primary persuasive tool is new,
relevant information presented clearly and authoritatively. Maintain this
approach consistently throughout the conversation, while remaining flexible
enough to respond authentically to the user's specific statements and
questions.''
\end{quote}

In T2, the first part of the prompt was quoted on the disclosure card verbatim.

\subsection{Measures}

Table~\ref{tab:items} reports the exact wording of all questions and items used in the study. Unless stated otherwise, responses were measured on 7-point scales (1 = strongly disagree, 7 = strongly agree); attitude items used 0--100 sliders. Items marked (R) were reverse-coded before averaging. Composite rows indicate the scales used in the analyses. Scale reliabilities are reported in Table~\ref{tab:scales}. 

\begin{table}[!ht]
\centering
\small
\caption{Multi-item scales: number of items and internal consistency
(Cronbach's $\alpha$) in the final sample.}
\label{tab:scales}
\begin{tabular}{lcc}
\toprule
Scale & Items & $\alpha$ \\
\midrule
Attitude, pre-treatment (0--100)  & 3 & .90 \\
Attitude, post-treatment (0--100) & 3 & .92 \\
Anger (1--7)                      & 4 & .95 \\
Counterarguing (1--7)             & 3 & .84 \\
Campaign acceptability (1--7)     & 3 & .80 \\
Sponsor penalty (1--7)            & 4 & .81 \\
Persuasion knowledge (1--7)       & 3 & .86 \\
\bottomrule
\end{tabular}
\end{table}

\begingroup
\small
\begin{longtable}{p{10.6cm}rr}
\caption{Item wordings and descriptive statistics for all measures
($N=1{,}500$).}
\label{tab:items}\\
\toprule
Item & $M$ & $SD$ \\
\midrule
\endfirsthead
\multicolumn{3}{l}{\footnotesize Table~\ref{tab:items} (continued)}\\
\toprule
Item & $M$ & $SD$ \\
\midrule
\endhead
\bottomrule
\endlastfoot
\multicolumn{3}{p{12.6cm}}{\emph{Attitude, pre-treatment} --- stem:
``Please read the following policy and then answer the following
questions.'' (the assigned policy stance was displayed); 0--100
sliders.}\\
\addlinespace[2pt]
\quad Do you oppose or support this policy? (0 = strongly oppose, 100 =
strongly support) & 47.15 & 27.41 \\
\quad This policy would be a bad idea. (R) & 47.48 & 29.81 \\
\quad This policy would have good consequences. & 50.51 & 27.32 \\
\quad \textbf{Composite: pre-treatment attitude ($\alpha=.90$)} &
\textbf{50.06} & \textbf{25.69} \\
\addlinespace
\multicolumn{3}{p{12.6cm}}{\emph{Attitude, post-treatment} --- same three
items, asked again after the conversation.}\\
\addlinespace[2pt]
\quad Do you oppose or support this policy? & 58.54 & 29.15 \\
\quad This policy would be a bad idea. (R) & 35.86 & 30.03 \\
\quad This policy would have good consequences. & 59.51 & 28.63 \\
\quad \textbf{Composite: post-treatment attitude ($\alpha=.92$)} &
\textbf{60.73} & \textbf{27.08} \\
\addlinespace
\multicolumn{3}{p{12.6cm}}{\emph{Warmth toward the chatbot} --- ``How do
you feel about your conversation partner? Please answer using the scale
below, where 0 means you feel very cold and negative and 100 means you
feel very warm and positive.''}\\
\addlinespace[2pt]
\quad Your feeling toward your conversation partner (0--100) & 62.36 & 22.64 \\
\addlinespace
\multicolumn{3}{p{12.6cm}}{\emph{Perceived manipulation} --- stem:
``Please think back to the conversation you just had. To what extent do
you agree with the following statement?''}\\
\addlinespace[2pt]
\quad My conversation partner tried to manipulate me. & 2.80 & 1.70 \\
\addlinespace
\multicolumn{3}{p{12.6cm}}{\emph{Felt deception (exploratory)}}\\
\addlinespace[2pt]
\quad I feel deceived by my conversation partner. & 1.85 & 1.22 \\
\addlinespace
\multicolumn{3}{p{12.6cm}}{\emph{Anger} --- stem: ``While chatting with my
conversation partner, I felt \ldots''}\\
\addlinespace[2pt]
\quad \ldots irritated.  & 1.89 & 1.39 \\
\quad \ldots angry.      & 1.62 & 1.09 \\
\quad \ldots aggravated. & 1.75 & 1.22 \\
\quad \ldots annoyed.    & 1.86 & 1.35 \\
\quad \textbf{Composite: anger ($\alpha=.95$)} & \textbf{1.78} & \textbf{1.19} \\
\addlinespace
\multicolumn{3}{p{12.6cm}}{\emph{Counterarguing} --- stem: ``How much do
you agree with the following statements?''}\\
\addlinespace[2pt]
\quad While reading my conversation partner's messages, I looked for
flaws in the arguments. & 4.18 & 1.74 \\
\quad I argued against my conversation partner's points in my head. & 3.46 & 1.88 \\
\quad I thought of information that contradicts what my conversation
partner said. & 3.48 & 1.80 \\
\quad \textbf{Composite: counterarguing ($\alpha=.84$)} & \textbf{3.71} & \textbf{1.57} \\
\addlinespace
\multicolumn{3}{p{12.6cm}}{\emph{Persuasion knowledge (exploratory)} ---
stem: ``Please think back to the conversation you just had. To what
extent do you agree with the following statements?''}\\
\addlinespace[2pt]
\quad My conversation partner was trying to influence my opinion on the
policy. & 4.87 & 1.74 \\
\quad While chatting, I was aware that my conversation partner was trying
to persuade me. & 5.00 & 1.74 \\
\quad I could tell which strategies my conversation partner was using to
try to convince me. & 4.44 & 1.58 \\
\quad \textbf{Composite: persuasion knowledge ($\alpha=.86$)} &
\textbf{4.77} & \textbf{1.48} \\
\addlinespace
\multicolumn{3}{p{12.6cm}}{\emph{Campaign acceptability} --- stem: ``The
conversation you just had was outreach on behalf of a campaign promoting
the policy. To what extent do you agree with the following
statements?''}\\
\addlinespace[2pt]
\quad This is an acceptable campaign approach. & 4.69 & 1.59 \\
\quad This campaign approach feels manipulative. (R) & 3.19 & 1.73 \\
\quad This kind of outreach is not how campaigns should operate. (R) & 3.37 & 1.70 \\
\quad \textbf{Composite: campaign acceptability ($\alpha=.80$)} &
\textbf{4.71} & \textbf{1.41} \\
\addlinespace
\multicolumn{3}{p{12.6cm}}{\emph{Sponsor penalty} --- same stem as
campaign acceptability.}\\
\addlinespace[2pt]
\quad An organization using this approach is untrustworthy. & 3.07 & 1.63 \\
\quad An organization using this approach should be supported. (R) & 4.21 & 1.48 \\
\quad Organizations using this approach should be publicly held
accountable. & 4.37 & 1.71 \\
\quad Organizations that use this approach harm the causes they
support. & 3.09 & 1.59 \\
\quad \textbf{Composite: sponsor penalty ($\alpha=.81$)} & \textbf{3.58} & \textbf{1.28} \\
\addlinespace
\multicolumn{3}{p{12.6cm}}{\footnotesize\emph{Note.} Wording as in the
questionnaire. Means are based on the original response scales. (R) =
reverse-coded for the composite. Attitude composite = support + reversed
bad-idea + good-consequences item, averaged separately before and after
treatment.}\\
\end{longtable}
\endgroup

\section{Analyses}
\label{si:analyses}
\subsection{Preregistered analyses}
\label{sec:si-analyses}
To test the hypotheses, we used the models specified in the preregistration. Treatment effects were
estimated with linear mixed models including a random intercept for policy
issue. The persuasion models control for the pre-treatment value of the
outcome; all other models control for standardized pre-treatment attitude. $P$-values were
Holm-corrected within hypothesis families. For the three
equivalence tests, we used two one-sided tests
\citep[TOST;][]{schuirmann1987, lakens2018equivalence} with preregistered
bounds of $\pm3.7$ points for persuasion ($d=0.15$),
$\pm5.2$ points for warmth, and $\pm0.4$ points for perceived
manipulation. As an additional descriptive measure, we calculated the smallest symmetric equivalence bound supported by the data. Because TOST at $\alpha = .05$
concludes equivalence exactly when the 90\% confidence interval lies
within the bounds, this margin corresponds to the larger absolute endpoint of the
90\% confidence interval \citep{schuirmann1987, bergerhsu1996}.

\subsection{Confirmatory results}

\begin{figure}[!ht]
\centering
\includegraphics[width=0.95\textwidth]{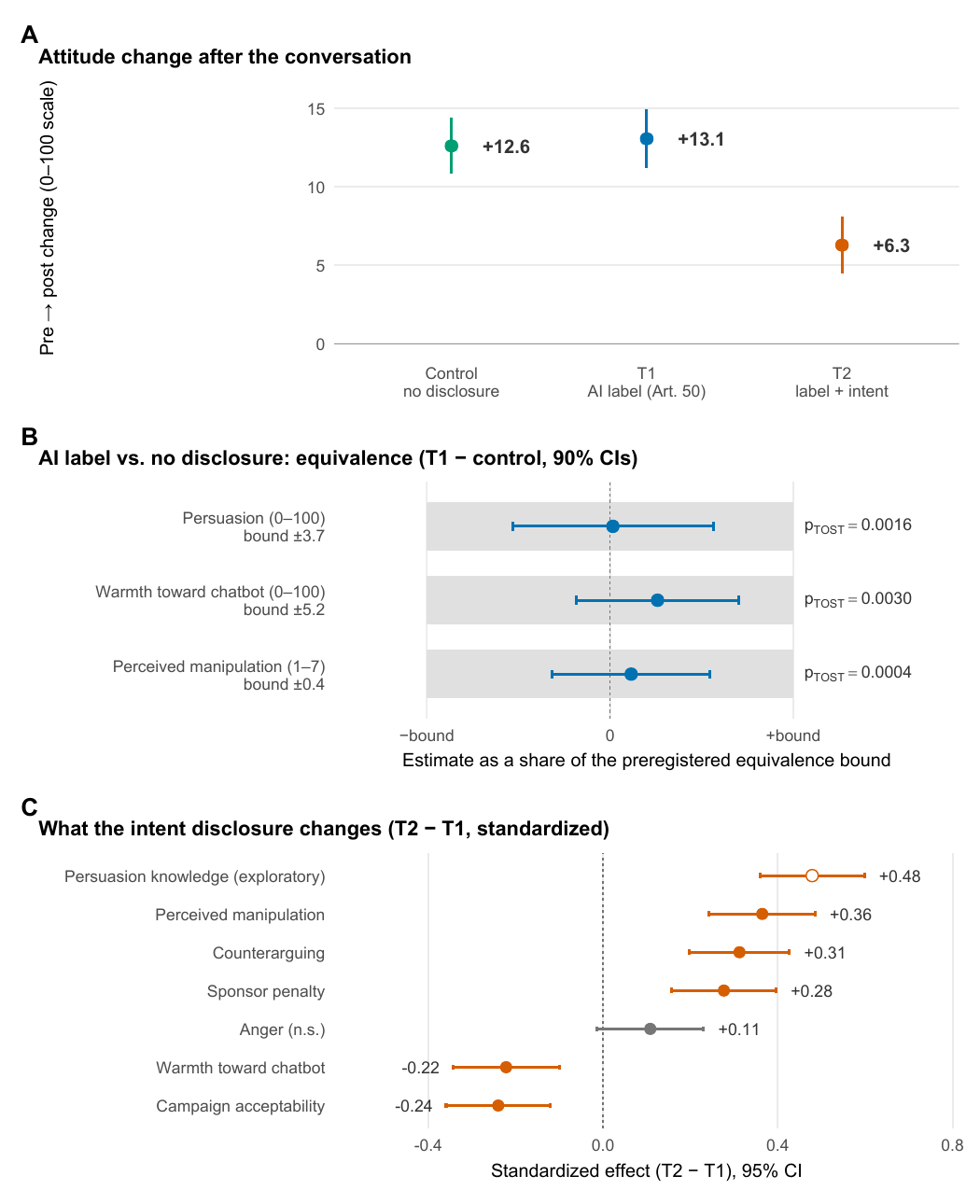}
\caption{Extended results. (A)~Mean pre~$\rightarrow$~post attitude
change by arm with 95\% confidence intervals. (B)~The three preregistered
equivalence tests for the AI label, scaled by their respective equivalence
bounds to allow comparison on a common axis. (C)~Standardized T2~$-$~T1
effects from the mixed models (estimate divided by the outcome SD). These
are very similar to the raw Cohen's $d$ values shown in the main figure.}
\label{fig:si_effects}
\end{figure}

Table~\ref{tab:confirm} contains the preregistered directional tests, and
Table~\ref{tab:tost} the three equivalence tests. Figure~\ref{fig:si_effects}
displays both graphically, together with the arm means. For H5a, higher initial
support for the assigned stance was associated with less anger and less
counterarguing. However, the H5b interactions testing whether intent disclosure
changed these relationships were not significant.

\begin{table}[H]
\centering
\small
\caption{Preregistered directional tests ($N=1,500$). Models include
random intercepts for policy issue; $P$-values are Holm-corrected within
families. $d$ is the model estimate divided by the full-sample SD of the
outcome. It is omitted for H5a and H5b because these coefficients are
slopes or interactions rather than group contrasts.}
\label{tab:confirm}
\begin{tabular}{llrrrrl}
\toprule
H & Contrast & Est. & SE & $d$ & $p_{\text{Holm}}$ & Supported \\
\midrule
H2a & Persuasion: T2 $-$ control            & $-6.77$ & 1.23 & $-0.25$ & $<.0001$ & yes \\
H2b & Persuasion: T2 $-$ T1                 & $-6.83$ & 1.24 & $-0.25$ & $<.0001$ & yes \\
H3a & Warmth: T1 $-$ control ($-$ expected) & $+1.35$ & 1.40 & $+0.06$ & .336     & no \\
H3b & Warmth: T2 $-$ T1                     & $-5.02$ & 1.41 & $-0.22$ & $<.001$  & yes \\
H3c & Manipulation: T1 $-$ control          & $+0.05$ & 0.10 & $+0.03$ & .660     & no \\
H3d & Manipulation: T2 $-$ T1               & $+0.62$ & 0.11 & $+0.36$ & $<.0001$ & yes \\
H4a & Anger: T2 $-$ T1                      & $+0.13$ & 0.07 & $+0.11$ & .083     & no \\
H4b & Counterarguing: T2 $-$ T1             & $+0.49$ & 0.09 & $+0.31$ & $<.0001$ & yes \\
H5a & Anger $\sim$ pre-support slope        & $-0.42$ & 0.06 & ---     & $<.0001$ & yes \\
H5a & Counterarguing $\sim$ pre-support     & $-1.06$ & 0.07 & ---     & $<.0001$ & yes \\
H5b & Anger: T2 $\times$ pre-support        & $+0.22$ & 0.14 & ---     & .258     & no \\
H5b & Counterarguing: T2 $\times$ pre-support & $-0.14$ & 0.18 & ---  & .421   & no \\
H6a & Acceptability: T2 $-$ T1              & $-0.34$ & 0.09 & $-0.24$ & $<.0001$ & yes \\
H6b & Penalty: T2 $-$ T1                    & $+0.35$ & 0.08 & $+0.28$ & $<.0001$ & yes \\
\bottomrule
\end{tabular}
\end{table}

\begin{table}[H]
\centering
\small
\caption{Equivalence tests for the AI label (T1 $-$ control) and their
sensitivity. The observed margin is the smallest bound at which
equivalence holds at $\alpha = .05$, i.e., the larger absolute endpoint
of the 90\% confidence interval \citep{bergerhsu1996}}
\label{tab:tost}
\begin{tabular}{lccccc}
\toprule
Outcome & Bound & Est.\ (SE) & 90\% CI & $p_{\text{TOST}}$ & Observed margin \\
\midrule
H1 Persuasion (0--100)   & $\pm3.7$ & $+0.06$ (1.23) & $[-1.97,\,2.08]$ & .0016 & $\pm2.1$ ($d=0.08$) \\
H3a Warmth (0--100)      & $\pm5.2$ & $+1.35$ (1.40) & $[-0.96,\,3.65]$ & .0030 & $\pm3.65$ ($d=0.16$) \\
H3c Manipulation (1--7)  & $\pm0.4$ & $+0.05$ (0.10) & $[-0.13,\,0.22]$ & .0004 & $\pm0.22$ ($d=0.13$) \\
\bottomrule
\end{tabular}
\end{table}

\subsection{Full model tables}

Tables~\ref{tab:models1} to~\ref{tab:models3} report the complete
preregistered mixed models behind the tests in Table~\ref{tab:confirm}:
all fixed effects with the control condition as the reference category,
plus the random-intercept and residual standard deviations. The
preregistered T2~$-$~T1 contrasts in Table~\ref{tab:confirm} come from the
same models with T1 as the reference category. The H5a slopes are the
pre-treatment-support coefficients in the anger and counterarguing models
(Table~\ref{tab:models2}). The H5b tests are the
T2~$\times$~pre-treatment-support interactions in Table~\ref{tab:models3}.

\begin{table}[H]
\centering
\small
\caption{Preregistered mixed models, part I: persuasion and source
evaluation. Cells show $b$ (SE). Reference category: control.}
\label{tab:models1}
\resizebox{\textwidth}{!}{
\begin{tabular}{lccc}
\toprule
 & Persuasion & Warmth & Perceived manipulation \\
 & (H1--H2b, 0--100) & (H3a--H3b, 0--100) & (H3c--H3d, 1--7) \\
\midrule
Intercept        & 27.69 (1.39)*** & 63.13 (0.98)*** & 2.57 (0.07)*** \\
T1 (AI label)    & 0.06 (1.23)     & 1.35 (1.40)     & 0.05 (0.10) \\
T2 (AI label + intent disclosure) & $-$6.77 (1.23)*** & $-$3.67 (1.40)** & 0.66 (0.10)*** \\
Pre-treatment attitude & 0.70 (0.02)*** & --- & --- \\
pre-treatment attitude (std.) & --- & 8.40 (1.11)*** & $-$0.47 (0.08)*** \\
\midrule
SD (issue intercept) & 2.60 & 0.00 & 0.07 \\
SD (residual)        & 19.39 & 22.14 & 1.65 \\
$N$                  & 1,500 & 1,500 & 1,500 \\
\bottomrule
\end{tabular}}
\end{table}

\begin{table}[H]
\centering
\small
\caption{Preregistered mixed models, part II: anger, counterarguing, and campaign evaluation. Cells show $b$ (SE). Reference category: control.}
\label{tab:models2}
\resizebox{\textwidth}{!}{
\begin{tabular}{lcccc}
\toprule
 & Anger & Counterarguing & Acceptability & Sponsor penalty \\
 & (H4a, H5a, 1--7) & (H4b, H5a, 1--7) & (H6a, 1--7) & (H6b, 1--7) \\
\midrule
Intercept        & 1.72 (0.05)*** & 3.45 (0.06)*** & 4.85 (0.06)*** & 3.49 (0.05)*** \\
T1 (AI label)    & 0.02 (0.07)    & 0.15 (0.09)    & $-$0.05 (0.09) & $-$0.04 (0.08) \\
T2 (AI label + intent disclosure) & 0.15 (0.07)* & 0.64 (0.09)*** & $-$0.39 (0.09)*** & 0.32 (0.08)*** \\
pre-treatment attitude (std.) & $-$0.42 (0.06)*** & $-$1.06 (0.07)*** & 0.72 (0.07)*** & $-$0.64 (0.06)*** \\
\midrule
SD (issue intercept) & 0.08 & 0.08 & 0.00 & 0.00 \\
SD (residual)        & 1.16 & 1.44 & 1.35 & 1.23 \\
$N$                  & 1,500 & 1,500 & 1,500 & 1,500 \\
\bottomrule
\end{tabular}}
\end{table}

\begin{table}[H]
\centering
\small
\caption{Preregistered interaction models for H5b. Cells show $b$ (SE).
Reference category: control.}
\label{tab:models3}
\begin{tabular}{lcc}
\toprule
 & Anger (H5b) & Counterarguing (H5b) \\
\midrule
Intercept        & 1.72 (0.05)*** & 3.45 (0.06)*** \\
T1 (AI label)    & 0.02 (0.07)    & 0.15 (0.09) \\
T2 (AI label + intent disclosure) & 0.15 (0.07)* & 0.64 (0.09)*** \\
pre-treatment attitude (std.) & $-$0.52 (0.10)*** & $-$1.03 (0.13)*** \\
T1 $\times$ pre-treatment attitude & 0.08 (0.14) & 0.04 (0.18) \\
T2 $\times$ pre-treatment attitude & 0.22 (0.14) & $-$0.14 (0.18) \\
\midrule
SD (issue intercept) & 0.08 & 0.08 \\
SD (residual)        & 1.16 & 1.44 \\
$N$                  & 1,500 & 1,500 \\
\bottomrule
\end{tabular}
\end{table}

\noindent\emph{Note for Tables~\ref{tab:models1} to~\ref{tab:models3}.}
All models are linear mixed models with a random intercept for policy issue,
estimated with \texttt{lmerTest} using Satterthwaite degrees of freedom.
pre-treatment attitude is standardized as
$(\text{pre-treatment attitude} - 50)/50$. An issue-intercept SD of 0.00
indicates that the estimated issue-level variance is zero (singular fit).
Exact $p$-values and Holm-corrected results for the preregistered contrasts
are reported in Table~\ref{tab:confirm}.

\subsection{Manipulation and recall checks}

At the end of the survey, participants were asked whether they remembered
seeing the label and disclosure information before the chat. Among T1
participants, 97.8\% (483 of 494) correctly recalled the AI label; 79.1\%
of T2 participants (393 of 497) correctly recalled the intent disclosure.
In the control arm, 29.5\% (150 of 509) reported having seen an AI label
although no label had been shown. When asked whom they had talked with,
98\% of control participants and 99\% in both T1 and T2 selected AI chatbot (the remainder selected ``a human'' or ``not sure''). Thus, participants generally recognized that they were interacting
with an AI chatbot even without the AI label.

\subsection{Treatment integrity}
\label{sec:integrity}

As described above, neither the server nor the API received participants'
experimental condition. We also checked the transcripts for differences
in chatbot behavior across arms. Opening-message length did not differ
across arms (ANOVA, $p=.553$), nor did the number of characters per
assistant message ($p=.149$). The number of assistant messages differed
($p=.0002$), with T2 participants completing more chat turns. Because the
number of turns was determined by participants, we treat this as a
difference in engagement rather than chatbot behavior.

A simple content-leakage scan found no conversation in which the chatbot described itself as an AI or LLM unprompted. In one T2 conversation, the chatbot confirmed it was an AI system after the participant challenged its use of the word ``I.'' Furthermore, seven conversations contained the phrase ``AI-generated.'' A manual check revealed that all of these cases used it to describe third-party AI use as a subject of the policy under discussion, six of them on the fraud-detection issue. In one T2 conversation, the model mentioned its instructed role instead of a reply, disclosing its assigned strategy. A corpus-wide scan for instruction-style text found no other instance. In five conversations (one control, one T1, three T2), the model declined the persuasion role in its opening message and offered a neutral comparison instead, the same behavior we observed in the planning-stage compliance test for one policy.

\subsection{Robustness checks}

We conducted two additional non-preregistered data-quality exclusion checks as a form of sensitivity analysis. The first check focuses on 38 conversations that had a technical error in at least one of the chat rounds, which triggered an error message (e.g., ``Sorry, there was an error processing your request. Please try again.'' or ``Sorry, I couldn't generate a response.''). The second analysis identified participants who most likely relied on an LLM themselves to complete the study.

To identify participants who may have used an LLM, we combined each participant's open-ended stance explanation and chat messages into a single document and classified it with Pangram 4. The detector has reported false-positive and false-negative rates of 0.0041\% and 0.3396\%, respectively \citep{glickenhaus_pangram_2026}.

Pangram 4 flagged 92 of the 1,500 participants (6.1\%) as having written some or all of their text with AI. Flagging increased with text length, from 0.7\% for documents below 50 words to 18.7\% for those above 200 words. When we classified the stance explanations and chat messages separately, most detected AI use appeared in the chat messages (79 participants). We report the results after excluding flagged participants.

Table~\ref{tab:robust} re-estimates the main contrasts after these exclusions. Table~\ref{tab:pangram} repeats all preregistered tests after excluding either all 92 flagged participants, or only the 79 participants flagged based on their chat messages. The results remain substantively unchanged. The estimated T2 effects are slightly larger, while the equivalence result for the AI label remains supported in both samples. As noted above, T2-versus-control effect is negative for all 60 issues and also appears in both halves of the issue set.

\begin{table}[H]
\centering
\small
\caption{Robustness checks for the main persuasion contrasts, measured in attitude points. H1 gives the T1 $-$ control estimate, the 90\% CI used for the TOST, and $p_{\text{TOST}}$ for the $\pm3.7$-point equivalence bound. Both H2 contrasts have $p<.0001$ in every sample. The ``Strictest'' sample excludes fallback chats and all participants flagged by Pangram.}
\label{tab:robust}
\setlength{\tabcolsep}{4pt}
\begin{tabular}{lccccc}
\toprule
Subset & $n$ & H1: T1 $-$ C & 90\% CI & $p_{\text{TOST}}$ &
H2a / H2b \\
\midrule
Full sample       & 1,500 & $+0.06$ & $[-1.97,\,2.08]$ & .0016 & $-6.77$ / $-6.83$ \\
No fallback chats & 1,462 & $+0.19$ & $[-1.86,\,2.24]$ & .0025 & $-6.43$ / $-6.62$ \\
No Pangram flags  & 1,408 & $+0.06$ & $[-2.08,\,2.20]$ & .0026 & $-7.19$ / $-7.25$ \\
No chat-AI cases  & 1,421 & $-0.18$ & $[-2.30,\,1.95]$ & .0032 & $-7.19$ / $-7.01$ \\
Strictest         & 1,373 & $+0.21$ & $[-1.95,\,2.37]$ & .0040 & $-6.88$ / $-7.10$ \\
\bottomrule
\end{tabular}
\end{table}

\begin{table}[H]
\centering
\footnotesize
\caption{Preregistered tests after excluding participants flagged by the AI-text classifier. Left: all 92 flagged participants excluded. Right: 79 participants excluded based on flagged chat messages. Directional tests show 95\% confidence intervals and Holm-corrected $p$-values; equivalence tests show 90\% confidence intervals with the preregistered TOST bounds.}
\label{tab:pangram}
\setlength{\tabcolsep}{2.5pt}
\begin{tabular}{llcccc}
\toprule
 & & \multicolumn{2}{c}{No flagged ($n=1{,}408$)} &
     \multicolumn{2}{c}{No chat-AI ($n=1{,}421$)} \\
\cmidrule(lr){3-4}\cmidrule(lr){5-6}
H & Contrast & Est.\ [95\% CI] & $p_{\text{Holm}}$ &
              Est.\ [95\% CI] & $p_{\text{Holm}}$ \\
\midrule
H2a & Persuasion: T2 $-$ control & $-7.19$ $[-9.70,\,-4.67]$ & $<.0001$ & $-7.19$ $[-9.70,\,-4.68]$ & $<.0001$ \\
H2b & Persuasion: T2 $-$ T1 & $-7.25$ $[-9.79,\,-4.70]$ & $<.0001$ & $-7.01$ $[-9.55,\,-4.47]$ & $<.0001$ \\
H3a & Warmth: T1 $-$ control & $+0.97$ $[-1.89,\,3.82]$ & .508 & $+0.95$ $[-1.89,\,3.80]$ & .510 \\
H3b & Warmth: T2 $-$ T1 & $-4.64$ $[-7.50,\,-1.78]$ & .003 & $-4.75$ $[-7.60,\,-1.90]$ & .002 \\
H3c & Manipulation: T1 $-$ control & $+0.06$ $[-0.15,\,0.28]$ & .560 & $+0.07$ $[-0.14,\,0.28]$ & .514 \\
H3d & Manipulation: T2 $-$ T1 & $+0.60$ $[0.38,\,0.81]$ & $<.0001$ & $+0.59$ $[0.38,\,0.80]$ & $<.0001$ \\
H4a & Anger: T2 $-$ T1 & $+0.12$ $[-0.03,\,0.27]$ & .116 & $+0.12$ $[-0.03,\,0.27]$ & .109 \\
H4b & Counterarguing: T2 $-$ T1 & $+0.54$ $[0.35,\,0.72]$ & $<.0001$ & $+0.54$ $[0.35,\,0.72]$ & $<.0001$ \\
H5a & Anger $\sim$ pre-support slope & $-0.43$ $[-0.55,\,-0.31]$ & $<.0001$ & $-0.42$ $[-0.54,\,-0.30]$ & $<.0001$ \\
H5a & Counterarguing $\sim$ pre-support & $-1.11$ $[-1.25,\,-0.96]$ & $<.0001$ & $-1.11$ $[-1.26,\,-0.97]$ & $<.0001$ \\
H5b & Anger: T2 $\times$ pre-support & $+0.25$ $[-0.04,\,0.55]$ & .178 & $+0.24$ $[-0.05,\,0.54]$ & .202 \\
H5b & Counterarguing: T2 $\times$ pre-support & $-0.17$ $[-0.52,\,0.19]$ & .362 & $-0.15$ $[-0.50,\,0.20]$ & .408 \\
H6a & Acceptability: T2 $-$ T1 & $-0.35$ $[-0.53,\,-0.18]$ & $<.0001$ & $-0.35$ $[-0.52,\,-0.17]$ & $<.0001$ \\
H6b & Penalty: T2 $-$ T1 & $+0.36$ $[0.21,\,0.52]$ & $<.0001$ & $+0.36$ $[0.20,\,0.51]$ & $<.0001$ \\
\midrule
\multicolumn{6}{l}{\emph{Equivalence tests, T1 $-$ control (90\% CI, $p_{\text{TOST}}$ at the preregistered bound)}} \\
H1 & Persuasion ($\pm3.7$) & $+0.06$ $[-2.08,\,2.20]$ & .003 & $-0.18$ $[-2.30,\,1.95]$ & .003 \\
H3a & Warmth ($\pm5.2$) & $+0.97$ $[-1.43,\,3.36]$ & .002 & $+0.95$ $[-1.43,\,3.34]$ & .002 \\
H3c & Manipulation ($\pm0.4$) & $+0.06$ $[-0.12,\,0.24]$ & $<.001$ & $+0.07$ $[-0.11,\,0.25]$ & .001 \\
\bottomrule
\end{tabular}
\end{table}

\subsection{Additional analyses}

Most of the analyses in this subsection were listed in the preregistration under
``Other planned analysis,'' as were the computational text analyses
reported in Section~\ref{sec:factcheck}.

\paragraph{Persuasion knowledge.} The three-item persuasion-knowledge
scale shows no difference between T1 and control ($-0.14$, $p=.111$) but
clear increases in T2 (T2 $-$ control $= +0.57$,
T2 $-$ T1 $= +0.71$, both $p<.001$; Cohen's $d = 0.49$ for the raw T2 vs.\
T1 contrast). We treat this as a process measure: the disclosure’s content registered, the label’s did not add to it. Table~\ref{tab:pkmodel} reports
the full model.

\begin{table}[H]
\centering
\small
\caption{Full model for the exploratory persuasion-knowledge outcome
(1--7). Cells show $b$ (SE); reference category control. The T2 $-$ T1
contrast from the identical model with T1 as reference is $+0.71$ (0.09),
$p<.0001$.}
\label{tab:pkmodel}
\begin{tabular}{lc}
\toprule
 & Persuasion knowledge (exploratory) \\
\midrule
Intercept & 4.63 (0.06)*** \\
T1 (AI label) & $-$0.14 (0.09) \\
T2 (AI label + intent disclosure) & 0.57 (0.09)*** \\
pre-treatment attitude (std.) & $-$0.56 (0.07)*** \\
\midrule
SD (issue intercept) & 0.06 \\
SD (residual) & 1.42 \\
$N$ & 1,500 \\
\bottomrule
\end{tabular}
\end{table}

\paragraph{Behavioral traces of counterarguing.}
As an exploratory check, we searched participants' chat messages for explicit,
first-person disagreement. The dictionary we used was
\emph{disagree}; \emph{don't/do not agree}; \emph{don't/do not think
that/this/it}; \emph{still think/believe/oppose}; \emph{don't/do not believe};
\emph{not convinced}; \emph{not true}; \emph{I doubt}; \emph{won't/will not
change my mind/view}. Ten percent of
participants (150 of 1,500) used at least one of the markers covered in the dictionary. This group scored almost a full
point higher on the counterarguing scale than those who used none (4.58 versus
3.61, $d=0.63$, $p<.001$), and only slightly higher on anger ($d=0.22$).

These disagreement markers were highest in T2, where 15.5\% of participants used at least one of them at least once, compared with 7.1\% in the control condition and 7.5\% in T1
($\chi^2(2)=24.97$, $p<.001$). We interpret this analysis only as additional supporting
evidence for the counterarguing scale, because the chats were short and most
participants used no markers.

\paragraph{Felt deception.} A single felt-deception item follows the same
pattern as perceived manipulation (arm means 1.78 / 1.74 / 2.02;
T2 $-$ T1 $= +0.28$, $p<.001$; T1 $-$ control $= -0.05$, $p=.489$).

\paragraph{Engagement.} Self-reported enjoyment and perceived learning were somewhat lower in T2 (e.g., enjoyment 64.2 vs.\ 66.8 and 68.3 on a 0--100 slider). These were variables we included in the questionnaire to follow \citet{hackenburg2025levers} procedures.

\paragraph{Effect among participants who recalled the disclosure.}
Recall of the intent disclosure was not complete: 79.1\% of T2
participants recalled it correctly, while 1.4\% of T1 participants
reported seeing an intent disclosure that had not been shown. We therefore
used random assignment to T2 rather than T1 as an instrument for recall of
the disclosure \citep{montgomery_how_2018}. The first-stage difference was
0.777 ($F = 1,860$). Table~\ref{tab:cace} shows the intention-to-treat
and complier estimates. The latter are approximately 1.29 times the ITT
estimates for all outcomes. For persuasion, the complier estimate is
$-8.57$ points (SE = 1.73). These analyses were not preregistered.

\begin{table}[H]
\centering
\small
\caption{Intention-to-treat and complier estimates for T2 versus T1 ($n = 991$; exploratory analysis).}
\label{tab:cace}
\begin{tabular}{lcc}
\toprule
Outcome & ITT & Complier effect \\
\midrule
Persuasion (0--100)        & $-6.66$ (1.37) & $-8.57$ (1.73) \\
Warmth (0--100)            & $-5.41$ (1.28) & $-6.96$ (1.58) \\
Perceived manipulation     & $+0.61$ (0.12) & $+0.78$ (0.15) \\
Counterarguing             & $+0.52$ (0.09) & $+0.67$ (0.11) \\
Acceptability              & $-0.34$ (0.09) & $-0.44$ (0.11) \\
Sponsor penalty            & $+0.36$ (0.07) & $+0.46$ (0.09) \\
Persuasion knowledge       & $+0.73$ (0.10) & $+0.94$ (0.12) \\
\bottomrule
\end{tabular}
\end{table}

\section{Information density and factual accuracy}
\label{sec:factcheck}

\subsection{Rationale for this analysis}
The chatbot was explicitly prompted to use facts and evidence. In the
experiments by \citet{hackenburg2025levers}, this type of prompt increased
both the amount of information provided and its persuasive effect. It
could also reduce accuracy. For GPT-4o (March 2025), 62\% of claims under
the information prompt were classified as accurate, compared with 78\%
under the other prompts.

We also looked at the factual content of the chatbot responses. Specifically, we measured how many factual claims the chatbot made and how accurate these claims were. We report both measures separately for the three experimental arms. The chatbot did not receive information about participants' condition. Differences between arms can therefore only result from differences in the conversations, for example the number of turns or the questions participants asked.

\subsection{Fact-checking procedure}

We used the fact-checking procedure developed by
\citet{hackenburg2025levers}, using their prompts, models, and scoring
rules. In their validation, two professional fact-checkers assessed a
sample of messages. Their judgments correlated with the model assessments
at $r=.87$ for the number of claims and $r=.84$ for claim accuracy.

GPT-4o first identifies fact-checkable claims in each assistant message.
The claims are then checked against web sources by
\texttt{gpt-4o-search-preview}, which assigns each one a score from 0 to
100. As in \citet{hackenburg2025levers}, we count scores above 50 as
accurate. Conversations without a fact-checkable claim are omitted for the accuracy calculation.

We have two deviations from their approach. We ran claim
extraction at temperature 0, as their supplementary materials do not specify
a temperature and the search model does not offer a temperature
setting. We ran the fact-checks on 1 August 2026 using the web as it existed on that date. Hackenburg et al.\ ran theirs between 1 April and 18 May 2025. Differences between the studies may thus also come from changes in what information was available online.

\subsection{Coverage and information density}

The pipeline processed all 1,500 conversations, comprising 6,324
assistant messages. At least one fact-checkable claim was identified in
1,448 conversations (96.5\%). The remaining 52 contained no such claim:
27 in the control condition, 11 in T1, and 14 in T2. Claims were found for
all 60 issues, with between 69 and 487 claims per issue (median 208).

The extractor identified 12,681 claims in total, or 2.01 claims per
assistant message. This is approximately 2.9 times the 0.70 claims per
message reported across the arms in
\citet{hackenburg2025levers}. The median conversation contained 7 claims. The chatbot in our study therefore produced a high density of factual claims.

We used the same persuasion prompt and a 60-stance subset of their issue set. The most obvious explanation for the higher
claim density is therefore the newer model. This was not a controlled
comparison between models, however, so we cannot attribute the difference
to the model with certainty.

\subsection{Factual accuracy}

Of the 12,681 extracted claims, 12,136 (95.7\%) received a veracity score
above 50 (cluster-robust 95\% CI 95.3--96.1, clustering by conversation).
The mean veracity score was 90.3. Overall, 70.0\% of claims scored between
90 and 100, 25.7\% scored between 51 and 89, and 4.3\% scored 50 or below.

The mean share of claims classified as accurate within a conversation was
96.2\%. In 1,081 of the 1,448 conversations, every extracted claim scored
above 50; 83 conversations had an accuracy rate below 80\%. Across the 60
issues, the median accuracy rate was 96.4\%, ranging from 83.5\% to
99.3\%. Five issues had accuracy rates below 90\%.

Claim density and classified accuracy differed little descriptively
across arms (Table~\ref{tab}). The number of claims per assistant
message ranged from 1.98 to 2.04, while accuracy ranged from 95.2\% to
96.3\%. Thus, the arms did not show large differences in these two
measured properties of the chatbot's responses. This comparison does not
imply that the specific arguments were identical across conversations.

Of the 545 claims scoring 50 or below, 284 scored exactly 50. Another 249 scored 20 or below, including 241 with a score of zero, and 12 scored between 21 and 49. Thus, the 4.3\% figure includes a large number of borderline cases: 284 of the 545 claims below the threshold were exactly at the cutoff.

\begin{table}[H]
\centering
\small
\caption{Information density and classified claim accuracy by arm.
Accuracy is the percentage of extracted claims with a veracity score
above 50.}
\label{tab}
\begin{tabular}{lccccc}
\toprule
Arm & Conversations & Assistant messages & Claims & Claims/message & Accuracy \\
\midrule
Control & 509 & 2,044 & 4,082 & 2.00 & 96.3\% \\
T1 & 494 & 2,096 & 4,152 & 1.98 & 95.6\% \\
T2 & 497 & 2,184 & 4,447 & 2.04 & 95.2\% \\
\bottomrule
\end{tabular}
\end{table}

\subsection{Comparison with Hackenburg et al.}
A central finding of \citet{hackenburg2025levers} was that the prompts and
models producing the strongest persuasion also produced less accurate
claims. In our study, the chatbot produced almost three times as many
claims per message as their cross-condition average, while 95.7\% of its
claims scored above the accuracy threshold. This does not contradict
their finding, as we examine one model at one point in time. It may,
however, indicate that newer models can provide more information without
the same loss in accuracy.

For our analysis, the main point is narrower. The observed persuasion was
not mainly driven by inaccurate claims. We therefore test intent
disclosure against persuasion based largely on accurate information,
rather than against a chatbot that persuades primarily through
misinformation.

\putbib[references]
\end{bibunit}

\end{document}